# PepSIRF: a flexible and comprehensive tool for the analysis of data from highly-multiplexed DNA-barcoded peptide assays

Zane W. Fink[1], Vidal Martinez[1], John Altin[2], and Jason T. Ladner[1]*

[1]The Pathogen and Microbiome Institute, Northern Arizona University, Flagstaff, AZ 86011
[2]TGen, Flagstaff, AZ 86005

*Corresponding author: jason.ladner@nau.edu

By coupling peptides with DNA tags (i.e., "barcodes"), it is now possible to harness high-throughput sequencing (HTS) technologies to enable highly multiplexed peptide-based assays, which have a variety of potential applications including broad characterization of the epitopes recognized by antibodies[1–3]. While the processing of HTS data, in general, is already well supported, there are very few software tools that have been developed for working with data generated in these highly-multiplexed peptide assays. In order to fill this gap, we present PepSIRF (Peptide-based Serological Immune Response Framework), which is a flexible and comprehensive software package designed specifically for the analysis of HTS data from highly-multiplexed peptide-based assays.

## General Design

PepSIRF is an open source command line, module-based program written in C++, which is available on GitHub (https://github.com/LadnerLab/PepSIRF). The current release (v1.3.1) includes 10 individual modules, which together support analysis starting with unprocessed HTS fastq files all the way to the identification of enriched peptides. We have also included a module designed for the deconvolution of highly-multiplexed viral peptide data, in order to conservatively predict past viral exposure events. **Figure 1** depicts a common workflow through the different modules. Below, we have included general descriptions for each of the PepSIRF modules. For detailed usage information see https://ladnerlab.github.io/PepSIRF.

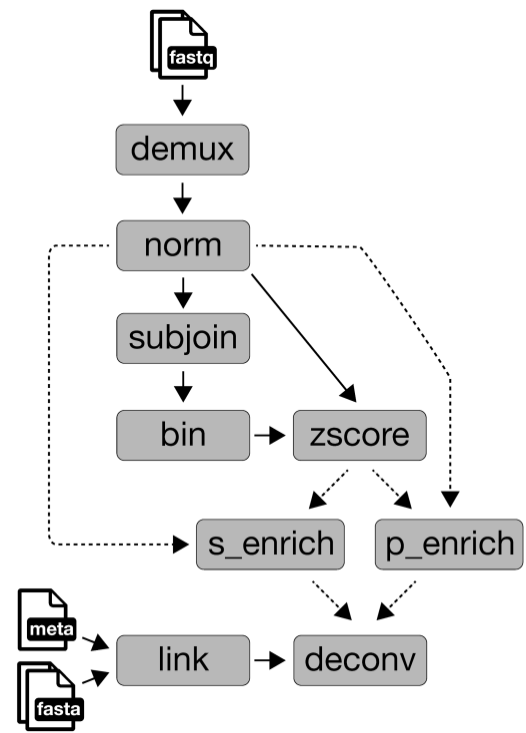

**Figure 1.** Graphical depiction of a common workflow through the nine analysis modules. Each grey box represents a separate PepSIRF module. An arrow between modules indicates a direct connection, with an output file from the upstream module being used as an input file for the downstream module. The dashed arrows represent alternative, parallel paths depending on the availability of replicates. Black and white icons represent a subset of the critical external files that must be provided.

# Modules

## Demux

The *demux* module is used to simultaneously demultiplex HTS data (i.e., assign reads to individual samples) and generate per sample counts for each peptide contained in the assay. Required inputs to this module are: 1) a fastq file (gzipped or decompressed), 2) a sample sheet containing sample names and associated index sequence names, 3) a fasta formatted file containing all of the possible index sequences and 4) a fasta formatted file containing the DNA tags for each peptide contained in the assay. A second fastq file can optionally be included to support a separate index read. To provide for flexibility in the arrangement of sequencing constructs, the positions and lengths of each index and the DNA tag are specified on the command line. The identification of each index and DNA tag proceeds through a stepwise search process until a match is found: 1) a perfect match to an expected sequence at the expected location, 2) a perfect match 1 nucleotide (nt) to the left of the expected position, 3) a perfect match 2 nt to the left of the expected position, 4) a perfect match 1 nt to the right of the expected position, 5) a perfect match 2 nt to the right of the expected position and 6) a match at the expected position with ≤ a user-specified number of mismatches. The output is a tab-delimited file containing read counts per peptide per sample, with one column for each sample and one row for each peptide contained in the assay. This module also provides support for libraries that contain multiple nucleotide encodings for the same peptide. In this case, two output files are generated, one with counts at the nucleotide-level and one with counts at the peptide-level.

## Info

The input and output files for many of the PepSIRF modules are tab-delimited matrix files with one column for each sample and one row for each peptide. The *info* module is a tool for quickly summarizing and extracting data from these matrix files. By default, the number of contained samples and peptides is simply printed to the screen. However, several optional flags can be used to extract sample names, peptide names and/or to generate sums for each column. This module is useful for understanding the content of a matrix file and also for generating input files for other modules.

## Subjoin

The *subjoin* module is used to generate new tab-delimited matrix files that represent subsets and/or combinations of existing tab-delimited matrix files. For this module, the user provides ≥1 tab-delimited matrix files, each with an associated list of sample names to include in the output file. The output will be a new tab-delimited matrix file containing the specified columns. If no sample list is provided for a given input file, all samples from that file are included in the output. The number of rows in the output file may exceed the number contained in the input files, if the input files do not contain identical lists of peptides. In this case, zeros will be added for samples when no associated value was present in the input file. By default, identical names from different input files are maintained as separate samples, with an input file-specific suffix added to the names to distinguish them in the output. However, with the "combine" option, this module can also be used to sum counts from multiple files (e.g., this would be appropriate for raw peptide counts from the same sample sequenced on two different runs). A common use case for this module is to create project-specific matrix files prior to downstream analysis.

**Norm**

The ***norm*** module is used to normalize raw read counts in order to account for differences in sequencing depth between samples. The input is a tab-delimited matrix containing non-normalized read counts; the output is an identical tab-delimited matrix containing normalized read counts. Two different normalization strategies are currently supported. The first is column-sum normalization for which the normalized read counts represent the number of reads mapped to a particular peptide per million total reads. The second method is equivalent to the size factors method described in Anders and Huber (2010)[4].

**Bin and Zscore**

The ***bin*** and ***zscore*** modules are used to calculate Z scores following the methodology laid out in Mina et al. (2019)[5]. First, the ***bin*** module can be used to generate groups of peptides with similar levels of abundance within negative control samples. The input for this module should be a tab-delimited matrix file containing normalized read counts from ≥1 negative control sample; the output is a tab-delimited file containing the peptide bins, with one bin per line. These bins can then be provided as input to the ***zscore*** module, along with a tab-delimited peptide count matrix, in order to calculate Z scores as a measurement of enrichment. Z scores are calculated separately for each individual sample and peptide, and these scores represent the number of standard deviations away from the mean, with mean and standard deviations calculated individually for each bin.

**S_enrich and P_enrich**

The ***s_enrich*** and ***p_enrich*** modules are used to generate lists of enriched peptides based on user-specified Z score and/or normalized read count thresholds. The user specifies the name of an output directory and within that directory, one file is generated per sample contained within the input matrices, as long as the sample contains at least one peptide that meets the enrichment thresholds. The generated output files are plain text and contain one peptide per line. Optionally, a raw, non-normalized read count matrix can also be provided and a minimum total sum of raw counts (i.e., total number of obtained reads for a sample) can be required for a sample to be processed. This option provides an easy way to filter out samples with low read counts. The ***s_enrich*** module assumes that only a single replicate assay was run for each sample, and therefore, each sample is considered individually. The ***p_enrich*** module, on the other hand, is for samples that were assayed in duplicate. For the ***p_enrich*** module, the user must also provide a tab-delimited file containing the names of the samples that should be treated as pairs, with one pair per line. The ***p_enrich*** module also requires that two thresholds are specified for both the Z scores and the normalized read counts. This allows the user to require different levels of enrichment from the two different replicates. At least one of the replicates must meet or exceed the higher threshold, but both replicates must meet or exceed the lower threshold. However, identical values can be provided for thresholds, if this behavior is not desired. Only one output file with enriched peptides is generated per pair.

**Link and Deconv**

The final two modules have been designed specifically for highly multiplexed analyses of antiviral antibody reactivity. However, the functionalities may also be useful in other contexts. The goal of these two modules is to predict the minimum list of viruses to which an individual has likely been exposed, based on the viral peptides that have been enriched through interaction with antibodies, while also considering shared sequence diversity among different viruses. The ***link*** module must first be used in order to generate a linkage file that relates individual peptides to viral taxa (e.g., species or strains). These connections are based on shared kmers between a peptide contained in the assay and a collection of protein sequences derived from various viral taxa. By default, a given taxon receives one point for each shared kmer. However, there is also an option to normalize these points by the relative abundance of the kmer in the total set of peptides contained in the assay. Inputs for the ***link*** module include 1) a fasta file containing the assay peptides, 2) a fasta file containing the complete collection of target protein sequences and 3) a tab-delimited metadata file that can be used to link target proteins to the taxonomic ID of interest; the output is a tab-delimited text file with two columns: 1) peptide name and 2) a comma-separated list of "taxon:score" for each taxon that shares ≥1 kmer with the assay peptide.

        The ***deconv*** module uses the output from the ***link*** module to identify the most parsimonious set of taxa that can explain a set of enriched peptides (see ***s_enrich/p_enrich***).
This is an iterative process. In each round, taxon-specific scores are generated by summing the taxon-level scores from each enriched peptide (contained in the output from the ***link*** module), and the taxon with the highest score is selected for inclusion in the output. All of the peptides that share ≥1 kmer with the selected taxon (i.e., peptides that may have been enriched through interaction with antibodies generated in response to the selected taxon) are then removed, and the process is repeated until no additional taxa meet a minimum user-specified score threshold. The ***deconv*** module also allows the user to specify criteria for multiple taxa to be considered as tied, if they result in similar scores and share kmers with an overlapping set of peptides. If multiple taxa are considered tied, then they will be reported together in a single line of the output file. The output of the ***deconv*** module is a tab-delimited text file with one line per selected taxon (or multi-taxa tie) and includes information about the number of enriched peptides that support the inclusion of the taxon, as well as the resulting score. The ***deconv*** module is designed to bulk process multiple independent lists of enriched peptides contained in the same directory, as would be created by the ***s_enrich*** or ***p_enrich*** modules. A separate output file will be generated for each input file.

**Acknowledgements**

Development of the software reported in this publication was supported by the State of Arizona Technology and Research Initiative Fund (TRIF), administered by the Arizona Board of Regents, through Northern Arizona University, and the National Institute On Minority Health And Health Disparities of the National Institutes of Health under Award Number U54MD012388. The content is solely the responsibility of the authors and does not necessarily represent the official views of the National Institutes of Health. ZWF was also supported by a Hooper Undergraduate Research Award and a Jean Shuler Research Mini-Grant.

**References**

1. Xu, G. J. *et al.* Viral immunology. Comprehensive serological profiling of human populations using a synthetic human virome. *Science* **348**, aaa0698 (2015).

2. Mohan, D. *et al.* PhIP-Seq characterization of serum antibodies using oligonucleotide-encoded peptidomes. *Nat. Protoc.* **13**, 1958–1978 (2018).

3. Kozlov, I. A. *et al.* A highly scalable peptide-based assay system for proteomics. *PLoS One* **7**, e37441 (2012).

4. Anders, S. & Huber, W. Differential expression analysis for sequence count data. *Nature Precedings* (2010) doi:10.1038/npre.2010.4282.1.

5. Mina, M. J. *et al.* Measles virus infection diminishes preexisting antibodies that offer protection from other pathogens. *Science* **366**, 599–606 (2019).